\documentclass[prd,%preprint,
,superscriptaddress,%showpacs,
nofootinbib,%
tightenlines ]{revtex4}
\usepackage{epsfig}
\usepackage[colorlinks=true,linkcolor=blue,urlcolor=blue,citecolor=blue]{hyperref}
\usepackage{dsfont}
\usepackage{amsmath}
\usepackage{amsfonts}
\usepackage{amssymb}
\usepackage{bm}

\newcommand{\ben}{\begin{displaymath}}
\newcommand{\een}{\end{displaymath}}
\newcommand{\be}{\begin{equation}}
\newcommand{\ee}{\end{equation}}
\newcommand{\bea}{\begin{eqnarray}}
\newcommand{\eea}{\end{eqnarray}}

\newcommand{\nn}{\nonumber \\ }

\begin{document}
%\preprint{MKPH-T-06-16}
\title{On the definition of the nucleon axial charge density % radius
}
\author{J.~Yu.~Panteleeva}
  \affiliation{Institut f\"ur Theoretische Physik II, Ruhr-Universit\"at Bochum,  D-44780 Bochum,
 Germany}
\author{E.~Epelbaum}
 \affiliation{Institut f\"ur Theoretische Physik II, Ruhr-Universit\"at Bochum,  D-44780 Bochum,
 Germany}
\author{J.~Gegelia}
 \affiliation{Institut f\"ur Theoretische Physik II, Ruhr-Universit\"at Bochum,  D-44780 Bochum,
 Germany}
\affiliation{Tbilisi State  University,  0186 Tbilisi,
 Georgia}
\author{U.-G.~Mei\ss ner}
 \affiliation{Helmholtz Institut f\"ur Strahlen- und Kernphysik and Bethe
   Center for Theoretical Physics, Universit\"at Bonn, D-53115 Bonn, Germany}
 \affiliation{Institute for Advanced Simulation (IAS-4), Forschungszentrum J\"ulich, D-52425 J\"ulich,
Germany}
\affiliation{Peng Huanwu Collaborative Center for Research and Education, Beihang University, Beijing 100191, China}
\affiliation{Tbilisi State  University,  0186 Tbilisi,
 Georgia}
 
\date{4 December 2024}
\begin{abstract}
We work out the spatial density distributions corresponding to the
axial-vector charge 
density operator for spin-1/2 systems using states
described by sharply localized wave packets. % in arbitrary Lorentz-frames.
The  static approximation, leading to the frequently assumed Breit-frame
distributions, is also considered.   
We discuss the interpretation of the resulting spatial densities
  in terms of the axial charge density. 
\end{abstract}

\maketitle

\section{Introduction and summary}
Following the seminal papers on
electron-proton scattering in the 60s
of the last century
\cite{Hofstadter:1958,Ernst:1960zza,Sachs:1962zzc} the three-dimensional  Fourier transform of the
charge form factor in the Breit frame is traditionally interpreted as the electric charge density
of the considered hadron.  
Analogous interpretation has also been suggested for the Fourier transforms of
the gravitational form factors and for various local distributions 
\cite{Polyakov:2002wz,Polyakov:2002yz,Polyakov:2018zvc}.   
On the other hand, in last two decades it has been repeatedly pointed out that 
the identification of spatial density
distributions with the Fourier transform of the corresponding form
factors in the Breit frame is problematic
\cite{Burkardt:2000za,Miller:2007uy,Miller:2009qu,Miller:2010nz,Jaffe:2020ebz,Miller:2018ybm,Freese:2021czn}.    
 
The issues of a proper definition and interpretation of the spatial density distributions of matrix elements of local
operators have attracted much attention in last years, see,
e.g.,~Refs.~\cite{Burkardt:2000za, Miller:2007uy, Miller:2009qu,
  Miller:2010nz,
  Guo:2021aik,Panteleeva:2021iip,Freese:2021mzg,Lorce:2020onh,Lorce:2022cle,Lorce:2018egm,
  Freese:2021czn,Jaffe:2020ebz,Miller:2018ybm}.  
A novel definition of the three-dimensional charge density using
sharply localized wave packet states has been suggested in
Ref.~\cite{Epelbaum:2022fjc}. 
In fact, similar results have been published long ago in largely unknown work of Ref.~\cite{Fleming:1974af}. 
In Refs.~\cite{Epelbaum:2022fjc,Panteleeva:2022khw,Panteleeva:2022uii,Alharazin:2022xvp,Panteleeva:2023evj} 
the three-dimensional spatial densities, possessing the usual probabilistic interpretation, have been defined in
 the zero average momentum frame (ZAMF)  for systems with various
 spins by using spherically symmetric sharply localized wave 
packets (here we deleted: "without using any approximations").\footnote{The ZAMF is defined as a Lorentz frame in
 which the mean value of the 
three-momentum for the state, specified by the given spherically symmetric packet, is zero. 
%For wave packets with a sharp localization around an eigenstate of the four-momentum operator, the ZAMF coincides with the rest-frame of the system.
} 
New definition has advantage compared to the Breit frame approach that it is also applicable to systems whose characteristic radii are of the order 
of (or less than) their Compton wavelengths.
By generalizing the definition to moving 
frames it was also shown that in the infinite-momentum frame (IMF), the new spatial densities turn
into the well-known two-dimensional distributions in the transverse
plane, multiplied by the delta-function in the longitudinal direction.

 Recently, the spatial densities of
 spin-$1/2$ systems corresponding to the axial-vector current have been studied in the 
 phase-space Wigner distributions approach \cite{Chen:2024oxx}. The
 authors of this paper 
 argued  that the axial charge
 density vanishes identically in the Breit frame and thus does not yield a
 nontrivial axial charge radius. For a review on the determinations of
 the axial radius of the nucleon, defined in terms of the slope of the
 axial-vector form factor, see Ref.~\cite{Bernard:2001rs}.

%\medskip

In this paper we work out the details of the novel definition
of the spatial distribution corresponding to the zeroth component of
the axial-vector current
for spin-$1/2$ systems. We build upon our earlier work
\cite{Epelbaum:2022fjc,Panteleeva:2022khw,Panteleeva:2022uii,Alharazin:2022xvp,Panteleeva:2023evj}
and employ sharply localized wave packet states to obtain the
corresponding spatial densities in the ZAMF as well as in IMF.\footnote{Calculations are similar and closely follow those of Ref.~\cite{Epelbaum:2022fjc}.} We also consider an alternative definition of the densities
by taking the static limit, which leads to the frequently used
Breit-frame distributions, and compare our results to the findings of
Ref.~\cite{Chen:2024oxx}. While the resulting spatial
distributions cannot be directly interpreted as axial charge densities, see
also Ref.~\cite{Chen:2024oxx}, we argue that
a proper definition of the axial charge densities follows by dropping
in the obtained expressions the overall spin-dependent factor that
does not encode information about the internal structure of the
system.  
Similarly to the electromagnetic case, the axial charge density defined via the Breit-frame distribution is
appropriate and, in fact, quite natural for heavy systems, however it is inappropriate for systems with radii of the order of, or smaller, than their Compton wave lengths. 
This motivates us to consider the ZAMF definition without relying on the static approximation
that corresponds to considering the system in states
  characterized by wave packets whose spatial extent is much larger
  than the Compton wavelength, yet much smaller than the
  corresponding radius \cite{Jaffe:2020ebz}. Spatial densities
  defined this way probe not only the internal structure of a system
  under consideration 
  but also the effects of delocalization. In contrast, using sharply
  localized wave packet states 
  allows one to
  reduce delocalization artifacts in the definition of spatial
  densities to, in principle, arbitrarily
  small amount.

Our paper is organized as follows. In sec.~\ref{sec2}, we apply the
framework developed in Refs.~\cite{Epelbaum:2022fjc,Panteleeva:2022khw,Panteleeva:2022uii,Alharazin:2022xvp,Panteleeva:2023evj} to work out the spatial
distributions corresponding to the zeroth component of the
axial-vector current in sharply-localized wave-packet states. The
considerations in this section are limited to the ZAMF. We also
consider here an alternative and more traditional approach that uses wave packets localized
at distances much larger than the Compton wave length of the system
which does, however, only provide wave-packet-independent
distributions in the static limit. In line with the observations
made in Ref.~\cite{Chen:2024oxx}, we find vanishing results for the
ZAMF densities in both approaches. To trace the origin of these
findings, the considerations of sec.~\ref{sec2} are generalized to
the spatial distributions in arbitrary moving frames in
sec.~\ref{sec3}. The interpretation of the vanishing distributions in
the ZAMF and a physically meaningful definition of the intrinsic
axial charge densities of a spin-1/2 system are provided in
sec.~\ref{sec4}.

% \medskip

\section{Spatial distributions in the ZAMF}
\label{sec2}

The normalization of the four-momentum eigenstates
$|p,s\rangle$ which we use to characterize the considered spin-$1/2$
system is specified by the equation 
\begin{equation}
\langle p',s'|p,s\rangle = 2 E (2\pi )^3 \delta_{s's}\delta^{(3)} ({\bf p'}-{\bf p})\,,
\label{NormStateN}
\end{equation}
where $p=(E,{\bf p})$, with $E=\sqrt{m^2+{\bf p}^2}$, and $m$ is the particle's mass. 

The matrix elements of  the axial-vector current operator 
between momentum eigenstates of a spin-$1/2$ system is parameterized in terms of two form factors
(utilizing  Lorentz invariance, isospin conservation and the discrete symmetries C, P and T, and
assuming the absence of second class currents):
\begin{eqnarray}
\langle p',s' | j^\mu_A ({\bf r} ,0)| p ,s\rangle & = & e^{ - i({\bf p'}-{\bf p})\cdot {\bf r}} \, \bar u(p',s') \left[ \gamma^\mu \gamma_5 G_A(q^2) 
+  \frac{ q^\mu \gamma_5}{2 M}\, G_P(q^2) \right] u(p,s) \,.
\label{eqMEN}
\end{eqnarray} 
Here, $p$ and $s$ ($p'$ and $s'$) denote the momentum and the polarization of the initial (final) states,
respectively, and the Dirac spinors are normalized as $ \bar u(p,s') u(p,s) = 2 m \delta_{s's}$.  
The momentum transfer is given as $q = p' - p$, while $G_A(q^2)$ and $G_P(q^2)$ are the axial and
induced pseudoscalar form factors, respectively. 
The factor $M$ introduced in front of the form factor $G_P$ is
commonly set equal to the particle's mass. However, as it mixes the
orders of the $1/m$ expansion used below, we distinguish between these
two quantities. In the final results we set $M=m$, meaning that these
mass scales are numerically equal.

%\medskip

To obtain local density distributions we calculate the matrix element
of the axial charge density
operator in a state localized in coordinate  space.  
For that purpose we consider a normalizable Heisenberg-picture state 
specified by a wave packet:
\begin{equation}
|\Phi, {\bf X},s \rangle = \int \frac{d^3 {p}}{\sqrt{2 E (2\pi)^3}}  \, \phi(s,{\bf p})
\, e^{-i {\bf p}\cdot{\bf X}} |p ,s \rangle\,. 
\label{statedefN2}
\end{equation}
From the normalization condition it follows that  
\begin{equation}
\int d^3 {p} \,  | \phi(s,{\bf p})|^2 =1\,.  
\label{normN}
\end{equation}
 The state $|\Phi, {\bf X},s \rangle$ above depends
  on the vector ${\bf X}$, whose interpretation will be given later.
To {\it define} spatial density distributions we employ spherically symmetric wave
packets with spin-independent profile functions  $\phi(s, {\bf p}) = \phi({\bf p}) = \phi(|{\bf p}|)$, which corresponds to ZAMF. 
This choice appears
legitimate due to the absence of a preferred direction in  space.
For convenience, we also introduce a dimensionless profile function
$\tilde \phi $ via 
\begin{equation}
\phi({\bf p}) = R^{3/2} \, \tilde \phi(R  {\bf p})\,,
\label{packageFormN}
\end{equation} 
where $R$ characterizes the size of the wave packet with small values
of $R$ corresponding to a sharp localization.

The matrix element of the axial charge density operator $j_A^0 $
  for the state defined in Eq.~(\ref{statedefN2}) is given by
\begin{eqnarray}
  j^0_{A\phi}({\bf r}) &\equiv &
                                 \langle \Phi, {\bf X},s' |
                                 %\hat
                                   j_A^0 ({\bf x}, 0 ) | \Phi, {\bf X},s
                                \rangle  \nn
                                &=&
                                \int \frac{d^3 {P} \, d^3 {q}}{(2\pi)^3 \sqrt{4 E E'}}\,
\, \bar u(p',s') \left[ \gamma^0 \gamma_5 G_A((E-E')^2- {\bf q}^2) +
                                    \frac{q^0 \gamma_5}{2 M} \,
                                    G_P((E-E')^2- {\bf q}^2) \right]
                                    u(p,s) \nn 
                &\times& \phi\bigg( {\bf P} -
\frac{\bf q}{2}\bigg) \, \phi^\star\bigg( {\bf P} +\frac{\bf q}{2}\bigg)  \, e^{ - i {\bf q}\cdot {\bf  r}} ,
\label{rhoint2N}
\end{eqnarray}
where ${\bf P}=({\bf p}' + {\bf p})/2$,   $E=\sqrt{m^2+ {\bf P}^2 - {\bf P}\cdot {\bf q} +{\bf q}^2/4 } $,
 $E'=\sqrt{m^2+ {\bf P}^2 + {\bf P}\cdot {\bf q} +{\bf q}^2/4 } $,
and we have introduced ${\bf  r} = {\bf  x}-{\bf  X}$. The spatial vector coordinates ${\bf  x}$,  ${\bf  X}$ and  ${\bf  r}$ are defined hier in ZAMF, while in the next section the same variables refer to moving frame.
  
Using the methods of Refs.~\cite{Gegelia:1994zz,Beneke:1997zp}, the
$R\to 0$ limit of the expression in Eq.~(\ref{rhoint2N}) can be
calculated  
without specifying the form factors and the profile function of the packet.
For $G_A\left( q^2\right)$ and $G_P\left( q^2\right)$ decreasing at  large $q^2$ faster than
$1/q^2$  and $1/(q^2)^2$, respectively, the result of $j^0_{A \phi
}({\bf r})$  in the $R\to 0$ limit is obtained using the method of
dimensional counting 
\cite{Gegelia:1994zz} by
substituting ${\bf P}= \tilde {\bf P}/R$, expanding the integrand
around $R=0$ and keeping terms up to the zeroth order.  Doing so we obtain
\begin{equation}
J^0_{A\phi}({\bf r})  =  \int \frac{d^3 {\tilde P} \, d^3 {q}}{(2\pi)^3} { {\bf \hat{\tilde P}}\cdot {\bm \sigma} } \,
 G_A\bigg[  \Big( {\bf {\hat{ \tilde P}}\cdot{\bf q} }\Big)^2 - {\bf q}^2\bigg] 
 \, |\tilde\phi({
  {\bf \tilde P}})|^2  \, e^{ - i {\bf q}\cdot {\bf  r}} 
  \,,
\label{rhoint3N}
\end{equation}
where ${\bm \sigma}$ is a three-vector of Pauli sigma matrices, ${\bf\hat{\tilde P}}= {\bf{\tilde P}}/\tilde {P}$ with $\tilde
{P} \equiv | {\bf \tilde  P}|$, and we use $J_A^0$ instead of $j_A^0$
for the density written as an operator in spin space.   
Using Eq.~(\ref{normN}), the integration over ${\bf \tilde P}$  for
spherically symmetric wave packets with $\tilde\phi( {\bf {\tilde 
  P}}) =  \tilde\phi(| {\bf {\tilde 
  P}}|) $ leads to 
\begin{equation}
J^0_{A,\text{ZAMF}}({\bf r}) = 0 \,.
\label{rhoint4N}
\end{equation}

%\medskip

Next, in order to obtain the axial charge distribution in the static limit, which we refer
to as ``naive'' following the terminology of
Ref.~\cite{Jaffe:2020ebz}, we expand the integrand in
Eq.~(\ref{rhoint2N}) in powers of $1/m$ and 
subsequently localize the wave packet by taking the limit $R \to 0$
\cite{Miller:2018ybm,Jaffe:2020ebz}.
The two leading-order contributions in the $1/m$-expansion of the integrand lead to 
the following expression:
\begin{equation}
J^0_{A \phi, \, \rm naive}({\bf r}) = \int \frac{d^3 {P} \, d^3 {q}}{(2\pi)^3} \left[ % \left\{ 
\frac{  {\bf P}\cdot {\bm \sigma} }{m} \, %\bar\chi_{s'}
 %\, \chi_s 
 G_A(- {\bf q}^2) - \frac{ { {\bf q}}\cdot {\bm \sigma} { {\bf P}}\cdot {\bf q}  }{4M m^2} %\, \chi_s 
\, G_P(- {\bf q}^2)  % + {\cal O}\left(\frac{1}{m^2}\right) 
\right]
               \,   \phi\bigg( {\bf P} -
\frac{\bf q}{2}\bigg) \, \phi^\star\bigg( {\bf P} +\frac{\bf q}{2}\bigg) \, e^{ - i {\bf q}\cdot {\bf  r}}  \,.
\label{rhoint2NRN}
\end{equation}
Also for the naive definition of the axial charge density in Eq.~(\ref{rhoint2NRN}), 
the $R\to 0$ limit of the expression in Eq.~(\ref{rhoint2NRN}) can be calculated by applying the method of dimensional counting
without specifying the form factors and the profile function
of the packet.
Again, for $G_A\left( q^2\right)$ and $G_P\left( q^2\right)$ decreasing at  large $q^2$-values faster than
$1/q^2$  and $1/(q^2)^2$, respectively, we obtain using the method of dimensional counting 
\cite{Gegelia:1994zz} 
\begin{eqnarray}
{ J}^0_{A ,\rm naive}({\bf r})  \; &\equiv& 0 \,,  
       \label{rhoint3RN}
\end{eqnarray}
where we have used the spherical symmetry of the wave
packet.\footnote{The  final result in Eq.~(\ref{rhoint3RN})  
 represents the spatial density for packets localized with $R$ much 
larger than the Compton wavelength $1/m$ yet  much smaller than any
other characteristic length scale of the system.  
For light hadrons, the characteristic length scales are comparable to the Compton
wavelength, therefore such a definition of spatial densities becomes
doubtful, see Ref.~\cite{Jaffe:2020ebz}, and references therein.}  

At first sight, the vanishing results of Eqs.~(\ref{rhoint3RN}) and
(\ref{rhoint4N}) seem to suggest that the Fourier transform of $G_A(q^2)$
does not correspond to the axial charge density of the considered
hadron, in agreement with the statements of Ref.~\cite{Chen:2024oxx}.  
To have a more detailed look at the problem we consider in the next
section our system from the point of view of a moving observer.

\section{Spatial distributions in moving frames}
\label{sec3}

To obtain the expressions of spatial density distributions in moving frames,
we follow the procedure of Ref.~\cite{Epelbaum:2022fjc} and replace the packet
 in Eq.~(\ref{rhoint2N})
  with its boosted expression. 
For that we express 
 a normalizable Heisenberg-picture state in a moving frame in terms of the spherically symmetric 
profile function $\phi ({\bf p})$ as (see also Ref.~\cite{Hoffmann:2018edo}):   
\begin{equation}
|\Phi, {\bf X},s \rangle _{\bf v}  =  \int \frac{d^3 {p}}{\sqrt{2 E
    (2\pi)^3}}  \, \sqrt{\gamma \Big( 1-\frac{{\bf v} \cdot {\bf 
                                        p}}{E} \Big)} \, \phi \Big(
                                        \Lambda_{\bf v}^{-1} {\bf p} \Big)
  \sum_{s_1}D_{s_1s}\bigg[ W\bigg( \Lambda_{\bf v}, 
    \frac{ \Lambda_{\bf v}^{-1} {\bf p} }{m} \bigg) \bigg]  e^{-i {\bf p\cdot X}} |p ,s_1 \rangle,  
\label{statedefN2Moving}
\end{equation} 
where  $\gamma = (1 - v^2)^{-1/2}$,
$E = \sqrt{m^2 + {\bf p}^2}$, $\Lambda_{\bf v}$ is the Lorentz boost
from the ZAMF to the moving frame, $\Lambda_{\bf v}^{-1} {\bf p} =   {\bf {\hat v}} \times ({\bf p}
\times   {\bf {\hat v}}) + \gamma ({\bf p} \cdot  {\bf {\hat v}} - v E
)  {\bf {\hat  v}}$, and $D_{s_1s}\left[ W\right]$ is a spin-1/2
representation of the 
corresponding Wigner rotation $W$  \cite{Weinberg:1995mt}.
In exact analogy to the case of the ZAMF, 
we find using the method of dimensional counting of Ref.~\cite{Gegelia:1994zz} (or the "strategy of regions" of Ref.~\cite{Beneke:1997zp}) that the only
non-vanishing contribution to spatial density distributions is
obtained by substituting ${\bf P} = {\bf \tilde  P}/R$, and expanding
around $R = 0$. Doing so for the moving frame we obtain 
\begin{align}
J^0_{A\phi,  {\bf v}}({\bf r}) & =  \int \frac{d^3 {\tilde P} \, d^3 {q}}{(2\pi)^3} 
\, \gamma \left(1  - {\bf v} \cdot {\bf\hat{\tilde  P}} \right) 
{ {\bf \hat{\tilde P}}\cdot {\bm \sigma} } \,  
 G_A\bigg[  \Big( { {\bf \hat{\tilde P}}\cdot{\bf q} }\Big)^2 - {\bf q}^2 \bigg]   
\, \big|\tilde\phi \big({\bf \tilde P}' \big)
\big|^2  \, e^{ - i {\bf q}\cdot {\bf  r}}\, ,
\label{rhoint3boostedN}
\end{align}
where we have introduced ${\bf \tilde P}' =
 {{\bf \hat  v}} \times \big( {\bf \tilde P} \times  {{\bf \hat  v}}\big)   +
  \gamma \big({\bf  \tilde  P}  \cdot  { {\bf \hat   v}}   -   v \tilde
  P\big)    {{\bf \hat  v}}$ and a unit vector ${\bf  {\hat{\tilde
        P}}'} \equiv {{\bf \hat  m}}$. 
Next, we change the integration variable ${\bf  \tilde P} \to {\bf
  \tilde P}'$ and define a vector-valued function 
\begin{equation}
  \label{Def_n}
  {\bf n} \big({\bf v},  { {\bf \hat m}} \big) = { {\bf \hat v}}
  \times \big( {\bf  \hat m} \times  { {\bf \hat v}}\big) 
+ \gamma \big( {\bf  \hat m} \cdot  { {\bf \hat v}}  + v )  {{\bf \hat v}} \,.
\end{equation}
Given that ${\bf \tilde P} =
 {{\bf \hat  v}} \times \big( {\bf \tilde  P}' \times  {{\bf \hat v}}\big)   +
  \gamma \big( {\bf \tilde  P}'  \cdot  { {\bf  \hat  v}}   +   v \tilde
  P'\big)    {{\bf \hat v}}$, it follows that  $ { {\bf \hat n}} = {\bf  {\hat {\tilde P}}} $.  
Taking into account the Jacobian of the change of variables ${\bf \tilde P} \to {\bf\tilde P}'$
we obtain from Eq.~(\ref{rhoint3boostedN}):
\begin{eqnarray}
  \label{rhoint3boosted2}
J^0_{A \phi,  {\bf v}}({\bf r}) & =  & \int  \frac{d {
                                     {\bf \hat m}} \,\tilde {P}'^2  d\tilde {P}' \, d^3
  {q}}{(2\pi)^3}
                                   \; \big|\tilde\phi \big(
                                     {\bf \tilde P} ' \big) \big|^2  \, e^{ - i {\bf q}\cdot {\bf  r}} \, %\bigg\{ 
                                  %   \bar\chi_{s'}
{ {\bf \hat{n}}\cdot {\bm \sigma} } \, %\chi_s
 %\left\{
 G_A \Big[ ({\bf \hat n}\cdot {\bf q}) ^2  - {\bf
                q}^2 \Big]  \,.
\end{eqnarray}
Using the spherical symmetry of $\tilde\phi \big(
{\bf \tilde P} ' \big) = \tilde\phi \big(|
{\bf \tilde P} '| \big) $ and Eq.~(\ref{normN}), the integration over
$\tilde P'$ can be easily carried out, and we obtain the final
expressions for the matrix element of the axial charge operator in moving frames:
\begin{eqnarray}
  \label{rhoint3boostedFinal}
J^0_{A,  {\bf v}}({\bf r}) & =  & \frac{1}{4 \pi}  \int  \frac{d {
                                     {\bf \hat m}} \, d^3
  {q}}{(2\pi)^3}
                                   \;  \, e^{ - i {\bf q}\cdot {\bf
                                  r}} \, { {\bf \hat{n}}\cdot {\bm
                                  \sigma} } \,  G_A \Big[ ({\bf \hat
                                  n}\cdot {\bf q}) ^2  - {\bf 
                q}^2 \Big]   \, .
\label{RhoCoordIndepNv}
\end{eqnarray}

The above result simplifies in two limiting 
cases. 
 First, for $v \to 0$ %and $\gamma = 1$, 
we have $  {\bf n} \big({\bf v},  {{\bf \hat m}} \big) = {\bf \hat 
  m} + {\bf v} +{\cal O}(v^2)$  and
    \begin{eqnarray}
J^0_{A,  {\bf v}}({\bf r}) & =  &  \frac{1}{4 \pi}   \int  \frac{d {
                                     {\bf \hat m}} \, d^3
  {q}}{(2\pi)^3}
                                   \;  \, e^{ - i {\bf q}\cdot {\bf
                                  r}} \, \left( \, {\bf {\hat m}\cdot {\bm
                                  \sigma} } + \, {\bf {v}\cdot {\bm
                                  \sigma} } \right) G_A \Big[ ({\bf \hat m}\cdot {\bf q}) ^2  - {\bf 
                q}^2 \Big] \nonumber\\
 & + &  \frac{1}{2 \pi}   \int  \frac{d {
                                     {\bf \hat m}} \, d^3
  {q}}{(2\pi)^3}
                                   \;  \, e^{ - i {\bf q}\cdot {\bf
                                  r}} \,  \, {\bf {\hat m}\cdot {\bm
                                  \sigma} }  \, {\bf {v}\cdot {\bm
                                  q} }  \, {\bf {\hat m}\cdot {\bm
                                  q} } \, G'_A \Big[ ({\bf \hat m}\cdot {\bf q}) ^2  - {\bf 
                q}^2 \Big]               
                +{\cal O}(v^2)  \, .
\label{RhoCoordIndepNvREDin}
\end{eqnarray}
For the integral over ${\bf \hat m}$ in the second line of Eq.~(\ref{RhoCoordIndepNvREDin}) we choose the $z$ axis in the direction of ${\bf q}$, and $x$ and $y$ axis 
such that ${\bf \sigma}$ lies in $xz$ plane. Introducing spherical coordinates ${\bf \hat m}=(\sin \theta \cos \phi, \sin \theta \sin \phi, \cos\theta)$  it is easily seen that the integral over $\phi$ leads to vanishing result. Next, by taking into account in the first line that 
the integration over an odd function of ${\bf \hat m}$ vanishes, we are left with
\begin{eqnarray}
  \label{rhoint3boostedFinal}
J^0_{A,  {\bf v}}({\bf r}) & =  &  \frac{1}{4 \pi}  \, {\bf {v}\cdot {\bm
                                  \sigma} } \int  \frac{d {
                                     {\bf \hat m}} \, d^3
  {q}}{(2\pi)^3}
                                   \;  \, e^{ - i {\bf q}\cdot {\bf
                                  r}} \, G_A \Big[ ({\bf \hat m}\cdot {\bf q}) ^2  - {\bf 
                q}^2 \Big] +{\cal O}(v^2)  \, .
\label{RhoCoordIndepNvRED}
\end{eqnarray}

The second limiting case corresponds to the IMF with $v \to 1$ and $\gamma
\to \infty$, for which the vector-valued function ${\bf \hat  n} $ turns to $
{\bf \hat v}$. In this case the integration over 
${\bf \hat  m}$ becomes trivial. Thus, for the matrix element of the axial charge density operator in the IMF we obtain  
\begin{eqnarray}
  \label{RhoIMF1}
  J^0_{A, \rm IMF} ({\bf r}) &= & { {\bf \hat{v}}\cdot {\bm \sigma} } 
   \int  \frac{d^3 {q}}{(2\pi)^3}
                                    \, e^{ - i {\bf q}\cdot {\bf  r}} \, %\bigg\{ 
                                  %   \bar\chi_{s'}
 %\chi_s
 %\left\{
 G_A \Big[ ({\bf \hat v}\cdot {\bf q}) ^2  - {\bf
                q}^2 \Big]  = \delta
(r_\parallel )  \, { {\bf \hat{v}}\cdot {\bm \sigma} }  \int  \frac{d^2 {q}_\perp}{(2\pi)^2}
                                    \, e^{ - i {\bf q}_\perp\cdot {\bf  r}_\perp} \, %\bigg\{ 
                                  %   \bar\chi_{s'}
%\chi_s
 %\left\{
 G_A \Big[   - {\bf
                q}_\perp^2 \Big] \,,
\end{eqnarray}
  where $r_\parallel \equiv |{\bf r}_\parallel |$ with ${\bf
    r}_\parallel = \bf r \cdot  {\bf \hat v} {\bf \hat v}$ and $r_\perp \equiv |{\bf r}_\perp |$ with ${\bf
    r}_\perp = \bf r - \bf r_\parallel $.
The appearance of  $\delta
(r_\parallel )$ in these expressions 
reflects the fact that in
the IMF, the system is Lorentz-contracted to a two-dimensional object
perpendicular to the velocity ${\bf \hat  v}$ of the moving frame.

To obtain the expressions of spatial density distributions for the
static approximation in moving frames, 
we apply the $1/m$ expansion to the integrand in Eq.~(\ref{rhoint3boostedN}) and obtain
at leading order: 
\begin{align}
J^0_{A\phi, {\rm naive}, {\bf v}}({\bf r}) & =  \int \frac{d^3 {P} \, d^3 {q}}{(2\pi)^3} 
\, \gamma \, 
\frac{ {\bf {P}}\cdot {\bm \sigma} }{m} \,
 G_A\left[- {\bf q}^2 \right]  
 \phi^* \big[ \Lambda_{\bf v}^{-1} \left( {\bf P} - {\bf q}/2 \right)
                                             \big]  \phi \big[
                                             \Lambda_{\bf v}^{-1}
                                             \left( {\bf P} + {\bf
                                             q}/2 \right) \big] 
\, e^{ - i {\bf q}\cdot {\bf  r}}\,, 
\label{rhoint3boostedNBF}
\end{align}
where  
 $\Lambda_{\bf v}^{-1} {\bf p} =   {\bf {\hat v}} \times ({\bf p}
\times   {\bf {\hat v}}) + \gamma ({\bf p} \cdot  {\bf {\hat v}} - m v)  {\bf {\hat  v}}  + {\cal O} (1/m) $.
Next, we change the integration variable ${\bf P} \to {\bf P}'$ by substituting ${\bf  P}' =
 {{\bf \hat  v}} \times \big( {\bf  P} \times  {{\bf \hat v}}\big)   +
  \gamma \big( {\bf P}  \cdot  { {\bf  \hat  v}}   -   m v \big)
  {{\bf \hat v}}$ and find that the Jacobian of the change of
  variables  
cancels the factor $\gamma$ in the integrand of
Eq.~(\ref{rhoint3boostedNBF}).  For sharply localized packets, the
integration over ${\bf P}'$ can be carried out easily by substituting
${\bf P}' = {\bf {\tilde P}}'/R$ and expanding at $R=0$. Doing so in
the leading order of the $1/m$ expansion we obtain: 
\begin{equation}
  \label{rhoint3boosted2}
J^0_{A,{\rm naive},  {\bf v}}({\bf r})  = 
{ {\bf {v}}\cdot {\bm \sigma} }   \int  \frac{ d^3
  {q}}{(2\pi)^3} \,
 G_A \left[  - {\bf
                q}^2 \right]  \, e^{ - i {\bf q}\cdot {\bf  r}}  \,.
\end{equation}

\section{Interpretation in terms of the axial charge density and discussion}
\label{sec4}

To shed light on the interpretation of the obtained spatial
distributions it is instructive to look at the corresponding
radial
moments in moving frames. In three spatial dimensions, we
find 
\bea
\label{Moments}
\langle r^{2k} \rangle_{\bf v} &=& \int d^3 r \, r^{2k} \, J^0_{A,
  {\bf v}}({\bf r})   = { {\bf {v}}\cdot {\bm \sigma} } \,
\big[ 2^{2k} k! G_A^{(k)} (0) \big] +{\cal O}(v^2) \,,
\eea
where $G_A^{(k)} (0)$ is the $k$th derivative of $G_A (q^2)$ with
respect to $q^2$ taken at $q^2 =0$.
% and the function $f (v)$ is given by
%\be
%f (v) = \frac{\gamma}{4 \pi v} \int d  {\bf \hat m} \, \frac{{\bf \hat
%    m} \cdot          {\bf \hat v} + v}{n}\,.
%\ee
%}
%Here, $n \equiv | {\bf n}|$ with the vector-valued function $\bf n$  defined in
%Eq.~(\ref{Def_n}).
%Notice that in the IMF, $f(v \to 1) = 1$.  
Similarly, for the naive distribution $J^0_{A,{\rm
    naive},  {\bf v}}({\bf r})$, one has   
\bea
\label{MomentsNaive}
\langle r^{2k} \rangle_{{\bf v}, \; \rm naive} &=& \int d^3 r \, r^{2k} \, J^0_{A,{\rm
    naive},  {\bf v}}({\bf r})   = { {\bf {v}}\cdot {\bm \sigma} } \,
\bigg[\frac{(2k +1)!}{k!} G_A^{(k)} (0) \bigg] \,,
\eea
Notice that the expressions in the square brackets of
Eqs.~(\ref{Moments}) and (\ref{MomentsNaive}) coincide with the
corresponding expressisons for the moments of
the electric charge distribution for a scalar particle of
Ref.~\cite{Epelbaum:2022fjc} upon replacing the form factor $G_A(q^2)$
with $F(q^2)$. 

We are now in the position to define the axial charge
density. Clearly and as already mentioned above, the distributions $J^0_{A, {\bf v}}({\bf r})$ and $J^0_{A,{\rm
    naive},  {\bf v}}({\bf r})$ cannot be interpreted as densities of
the axial charge. In particular, they change sign upon spatial
reflections as follows from their very definition in
Eq.~(\ref{rhoint2N}), i.e., they behave as pseudoscalar
densities.\footnote{This also explains the vanishing result in the
  ZAMF: The expressions must be proportional to the Pauli spin matrix,
  but because of the assumed symmetry of the wave packets, no vector
  is available in the ZAMF to make the densities rotationally
  invariant.} This also implies that the moments of these
distributions are pseudoscalar quantities, which is certainly not what one
expects from the density distribution of any quantity including the
axial charge. In particular, we expect the properly defined axial
charge density to
be normalized to the axial charge $g_A = G (0)$ in any frame. 
The expressions for the radial moments of the distributions $J^0_{A, {\bf v}}({\bf r})$ and $J^0_{A,{\rm
    naive},  {\bf v}}({\bf r})$ in Eqs.~(\ref{Moments}) and (\ref{MomentsNaive}) have the form of the spin- and
velocity-dependent prefactors, that do not depend on the
internal structure of the system, times the frame-independent factors
which incorporate the whole information about the axial charge distribution
encoded in the form factor $G_A$. This suggests the definition of the
properly normalized axial charge densities $\rho_{A, {\bf v}}({\bf r})$ and $\rho_{A,{\rm
    naive},  {\bf v}}({\bf r})$ by dropping the above-mentioned
structure-independent prefactors via
\be
J^0_{A, {\bf v}}({\bf r}) =:  { {\bf {v}}\cdot {\bm \sigma} } \,
 \rho_{A, {\bf v}}({\bf r}), \quad \quad
J^0_{A,{\rm
    naive},  {\bf v}}({\bf r}) =:  { {\bf {v}}\cdot {\bm \sigma} } \, \rho_{A,{\rm
    naive},  {\bf v}}({\bf r}). 
\label{defAD}
\ee
Notice that the normalization of densities defined via Eq.~(\ref{defAD}) is uniquely fixed by the condition  
that the axial charge of the considered system is given by $G_A(0)$.
This way, the naive definition of the axial charge density becomes
independent of the frame velocity and reduces to the  
traditional Breit-frame expression,
\begin{equation}
  \label{rhoint3boosted2rho}
\rho_{A,{\rm naive}}({\bf r})  =   \int  \frac{ d^3
  {q}}{(2\pi)^3} \, 
 G_A \left[  - {\bf
                q}^2 \right]  \, e^{ - i {\bf q}\cdot {\bf  r}}  \,.
\end{equation}
%while the result without invoking the static approximation takes the
%form 
%\be
%\rho_{A,{\bf v}}({\bf r})  =  \frac{\gamma}{4 \pi v \, f(v)}  \, \int  \frac{d {
%                                     {\bf \hat m}} \, d^3
%  {q}}{(2\pi)^3}
%                                   \;  \, e^{ - i {\bf q}\cdot {\bf
%                                       r}} \,
%\frac{{\bf \hat
%    m} \cdot          {\bf \hat v} + v}{n}\,
%                                   G_A \Big[ ({\bf \hat
%                                  n}\cdot {\bf q}) ^2  - {\bf 
%                q}^2 \Big]   \, .
%\label{Zexpr}
%\ee
On the other hand the density
in the ZAMF is obtained from Eq.~(\ref{RhoCoordIndepNvRED}) by dropping the factor ${\bf {v}}\cdot {\bm \sigma} $ and taking the limit $v\to 0$.  The resulting expression reads:
\begin{equation}
  \label{RhoIMF2rhoZ}
  \rho_{A, \rm ZAMF} ({\bf r}) \; = \;  \lim_{v \to 0}  \rho_{A, \bf
    v} ({\bf r}) \; = \; 
    \frac{1}{4 \pi} \int \frac{d^2 {\bf\hat m} \, d^3 {q}}{(2\pi)^3}
                                   \,   G_A \Big[ ({\bf \hat m}\cdot {\bf q}) ^2  - {\bf
                q}^2 \Big]  \, e^{ - i {\bf q}\cdot {\bf  r}}
%               \; 
%               = \;  \frac{1}{4 \pi}\int d^2 {\bf\hat v} 
%                                   \, \rho_{A, \rm IMF} ({\bf r}) 
                \,.
\end{equation}
Analogously in the infinite-momentum frame from Eq.~(\ref{RhoIMF1}) we obtain 
\begin{equation}
  \label{RhoIMF1rho}
  \rho_{A, \rm IMF} ({\bf r}) =  
   \int  \frac{d^3 {q}}{(2\pi)^3}
                                   \,   G_A \Big[ ({\bf \hat v}\cdot {\bf q}) ^2  - {\bf
                q}^2 \Big]  \, e^{ - i {\bf q}\cdot {\bf  r}}   = \delta
(r_\parallel ) \int  \frac{d^2 {q}_\perp}{(2\pi)^2}  \, 
 G_A \Big[   - {\bf
                q}_\perp^2 \Big] \, e^{ - i {\bf q}_\perp\cdot {\bf  r}_\perp}  \,.
\end{equation}
Equations~(\ref{RhoIMF1rho}) and (\ref{RhoIMF2rhoZ}) represent
our final result for the axial charge density in the
IMF and ZAMF, respectively, defined by sharply localized spherically
symmetric wave packets.\footnote{Notice that they do not depend on the induced pseudoscalar form factor $G_P$.}
%, while Eq.~(\ref{Zexpr}) generalizes these results to arbitrary moving frames. 
Interestingly, these
densities %in the ZAMF and IMF 
coincide with the
corresponding expressions for the electric charge density of Ref.~\cite{Epelbaum:2022fjc} upon replacing the electric form
factor $F(q^2)$ by the axial one $G_A (q^2)$. 
%This, however, does not apply to the densities in arbitrary moving frames (apart from the
%limiting cases of $v \to 0$ and $v \to 1$), for which the two results are different. 
The obtained expressions for the axial charge density are
applicable to systems regardless of their mass and, in particular,
remain valid for light systems, whose Compton wave length is comparable to or
larger than the characteristic scale(s) of the axial charge
distribution. 
%Notice further that the axial charge density in the ZAMF possesses the same holographic-like interpretation as discussed in Refs.~\cite{Epelbaum:2022fjc,Panteleeva:2022khw}. 
Last but not least, we emphasize that since the axial charge density
depends on the relative distance ${\bf r}= {\bf x} - {\bf X}$, ${\bf X}$ should be
interpreted as the position of the center of the axial charge
         distribution of the considered system.
        
We have also considered a naive definition
of the axial charge density obtained by taking the static limit
of the integrand in Eq.~(\ref{rhoint3boostedN}), which corresponds to 
localizing the system at distances much larger than the Compton
wavelength, while much smaller than other
characteristic scale(s) of the system. The resulting naive axial density
coincides with the three-dimensional Breit-frame distribution and is
frame-independent. Notice, however, that such a definition depends on
the form of the wave packet the system is prepared in if one
goes beyond the static approximation, see Ref.~\cite{Epelbaum:2022fjc} for more
details. Still, this way of defining the axial charge density is
appropriate and, in fact, quite natural for heavy systems. Notice,
however, that the sharp localization limit of $R \to 0$ does not
commute with the static limit, which is why the resulting (mass-independent)
expressions for the densities and their spherical moments in Eqs.~(\ref{Moments})
and (\ref{MomentsNaive}) differ from each other, see also
Ref.~\cite{Epelbaum:2022fjc}.
In particular, one has
\begin{equation}
\langle r^2\rangle_{A} = \int  d^3
  { r} \,  r^2  \, \rho_{A, \bf v} ({\bf r}) =4\,G_A'(0)\,,
\label{radsquareZAMF}
\end{equation}
while the conventional definition of the
axial charge radius is given by
\begin{equation}
\langle r^2\rangle_{A, \rm naive} = \int  d^3
  { r} \,  r^2  \, \rho_{A, \rm naive} ({\bf r}) =6\,G_A'(0)\,.
\label{radsquareBF}
\end{equation}

\acknowledgements
We thank C\'edric Lorc\'e for discussions and critical comments. 
This work was supported in part 
by the MKW NRW under the funding code NW21-024-A, by the Georgian
Shota Rustaveli National 
Science Foundation (Grant No. FR-23-856), 
by ERC  NuclearTheory (grant No. 885150) and ERC EXOTIC (grant No. 101018170),
by CAS through a President's International Fellowship Initiative (PIFI)
(Grant No. 2025PD0022),  and by the EU Horizon 2020 research and
innovation programme (STRONG-2020, grant agreement No. 824093).


\begin{references}

\bibitem{Hofstadter:1958}
R.~Hofstadter, F.~Bumiller, and M.~R.~Yearian,
Rev. Mod. Phys. {\bf 30}, 482 (1958).
%doi:10.1103/RevModPhys.30.482

\bibitem{Ernst:1960zza}
F.~J.~Ernst, R.~G.~Sachs and K.~C.~Wali,
%``Electromagnetic form factors of the nucleon,''
Phys. Rev. \textbf{119}, 1105-1114 (1960).
% doi:10.1103/PhysRev.119.1105
%199 citations counted in INSPIRE as of 11 Nov 2021

\bibitem{Sachs:1962zzc}
R.~G.~Sachs,
%``High-Energy Behavior of Nucleon Electromagnetic Form Factors,''
Phys. Rev. \textbf{126}, 2256-2260 (1962).
%doi:10.1103/PhysRev.126.2256
%269 citations counted in INSPIRE as of 11 Nov 2021

\bibitem{Polyakov:2002wz}
M.~V.~Polyakov and A.~G.~Shuvaev,
%``On'dual' parametrizations of generalized parton distributions,''
[arXiv:hep-ph/0207153 [hep-ph]].
%110 citations counted in INSPIRE as of 11 Nov 2021

\bibitem{Polyakov:2002yz} 
  M.~V.~Polyakov,
%  ``Generalized parton distributions and strong forces inside nucleons and nuclei,''
  Phys.\ Lett.\ B {\bf 555}, 57 (2003),
  %%doi:10.1016/S0370-2693(03)00036-4
  [hep-ph/0210165].

\bibitem{Polyakov:2018zvc}
M.~V.~Polyakov and P.~Schweitzer,
% ``Forces inside hadrons: pressure, surface tension, mechanical radius, and all that,''
Int.\ J.\ Mod.\ Phys.\ A \textbf{33} (2018) no.26, 1830025. 
%doi:10.1142/S0217751X18300259
%[arXiv:1805.06596 [hep-ph]].
%57 citations counted in INSPIRE as of 03 Apr 2020

\bibitem{Burkardt:2000za}
M.~Burkardt,
%``Impact parameter dependent parton distributions and off forward parton distributions for zeta ---\ensuremath{>} 0,''
Phys. Rev. D \textbf{62} (2000), 071503(R),
[erratum: Phys. Rev. D \textbf{66} (2002), 119903(E)],
%doi:10.1103/PhysRevD.62.071503
[arXiv:hep-ph/0005108 [hep-ph]].
%666 citations counted in INSPIRE as of 17 Feb 2021

\bibitem{Miller:2007uy}
G.~A.~Miller,
%``Charge Density of the Neutron,''
Phys. Rev. Lett. \textbf{99}, 112001 (2007),
%doi:10.1103/PhysRevLett.99.112001
[arXiv:0705.2409 [nucl-th]].
%235 citations counted in INSPIRE as of 11 Nov 2021

\bibitem{Miller:2009qu}
G.~A.~Miller,
%``Singular Charge Density at the Center of the Pion?,''
Phys. Rev. C \textbf{79}, 055204 (2009),
% doi:10.1103/PhysRevC.79.055204
[arXiv:0901.1117 [nucl-th]].
%35 citations counted in INSPIRE as of 11 Nov 2021

\bibitem{Miller:2010nz}
G.~A.~Miller,
%``Transverse Charge Densities,''
Ann. Rev. Nucl. Part. Sci. \textbf{60} (2010), 1-25, 
%doi:10.1146/annurev.nucl.012809.104508
[arXiv:1002.0355 [nucl-th]].
%100 citations counted in INSPIRE as of 16 Feb 2021

\bibitem{Jaffe:2020ebz}
R.~L.~Jaffe,
%``Ambiguities in the definition of local spatial densities in light hadrons,''
Phys. Rev. D \textbf{103} (2021) no.1, 016017,
%doi:10.1103/PhysRevD.103.016017
[arXiv:2010.15887 [hep-ph]].
%1 citations counted in INSPIRE as of 16 Feb 2021

\bibitem{Miller:2018ybm}
G.~A.~Miller,
%``Defining the proton radius: A unified treatment,''
Phys. Rev. C \textbf{99}, no.3, 035202 (2019),
%doi:10.1103/PhysRevC.99.035202
[arXiv:1812.02714 [nucl-th]].
%46 citations counted in INSPIRE as of 11 Nov 2021

\bibitem{Freese:2021czn}
A.~Freese and G.~A.~Miller,
%``Forces within hadrons on the light front,''
Phys. Rev. D \textbf{103}, 094023 (2021),
%doi:10.1103/PhysRevD.103.094023
[arXiv:2102.01683 [hep-ph]].


%\cite{Guo:2021aik}
\bibitem{Guo:2021aik}
Y.~Guo, X.~Ji and K.~Shiells,
%``Novel twist-three transverse-spin sum rule for the proton and related generalized parton distributions,''
Nucl. Phys. B \textbf{969}, 115440 (2021),
%doi:10.1016/j.nuclphysb.2021.115440
[arXiv:2101.05243 [hep-ph]].


%\cite{Panteleeva:2021iip}
\bibitem{Panteleeva:2021iip}
J.~Y.~Panteleeva and M.~V.~Polyakov,
%``Forces inside the nucleon on the light front from 3D Breit frame force distributions: Abel tomography case,''
Phys. Rev. D \textbf{104} (2021) no.1, 014008,
%doi:10.1103/PhysRevD.104.014008
[arXiv:2102.10902 [hep-ph]].

%\cite{Freese:2021mzg}
\bibitem{Freese:2021mzg}
A.~Freese and G.~A.~Miller,
%``Unified formalism for electromagnetic and gravitational probes: Densities,''
Phys. Rev. D \textbf{105}, no.1, 014003 (2022),
%doi:10.1103/PhysRevD.105.014003
[arXiv:2108.03301 [hep-ph]].

%\cite{Lorce:2020onh}
\bibitem{Lorce:2020onh}
C.~Lorc\'e,
%``Charge Distributions of Moving Nucleons,''
Phys. Rev. Lett. \textbf{125}, no.23, 232002 (2020),
%doi:10.1103/PhysRevLett.125.232002
[arXiv:2007.05318 [hep-ph]].
%16 citations counted in INSPIRE as of 09 Jan 2022

%\cite{Lorce:2022cle}
\bibitem{Lorce:2022cle}
C.~Lorc\'e, P.~Schweitzer and K.~Tezgin,
%``2D energy-momentum tensor distributions of nucleon in a large-Nc quark model from ultrarelativistic to nonrelativistic limit,''
Phys. Rev. D \textbf{106}, no.1, 014012 (2022),
%doi:10.1103/PhysRevD.106.014012
[arXiv:2202.01192 [hep-ph]].

%\cite{Lorce:2018egm}
\bibitem{Lorce:2018egm}
C.~Lorc\'e, H.~Moutarde and A.~P.~Trawi\'nski,
%``Revisiting the mechanical properties of the nucleon,''
Eur. Phys. J. C \textbf{79}, no.1, 89 (2019),
%doi:10.1140/epjc/s10052-019-6572-3
[arXiv:1810.09837 [hep-ph]].

%\cite{Epelbaum:2022fjc}
\bibitem{Epelbaum:2022fjc}
E.~Epelbaum, J.~Gegelia, N.~Lange, U.-G.~Mei\ss{}ner and M.~V.~Polyakov,
%``Definition of Local Spatial Densities in Hadrons,''
Phys. Rev. Lett. \textbf{129}, no.1, 012001 (2022),
%doi:10.1103/PhysRevLett.129.012001
[arXiv:2201.02565 [hep-ph]].

%\cite{Fleming:1974af}
\bibitem{Fleming:1974af}
%G.~N.~Fleming,
%%``Charge Distributions from Relativistic Form-Factors,''
%doi:10.1007/978-94-010-2274-3\_22
G.~N.~Fleming,  %Charge Distributions from Relativistic Form Factors. 
Physical Reality \& Math. Descrip., 357 (1974). 
% Springer, Dordrecht. https://doi.org/10.1007/978-94-010-2274-3\_22\,.

%\cite{Panteleeva:2022khw}
\bibitem{Panteleeva:2022khw}
J.~Y.~Panteleeva, E.~Epelbaum, J.~Gegelia and U.-G.~Mei\ss{}ner,
%``Definition of electromagnetic local spatial densities for composite spin-1/2 systems,''
Phys. Rev. D \textbf{106}, no.5, 056019 (2022),
%doi:10.1103/PhysRevD.106.056019
[arXiv:2205.15061 [hep-ph]].

%\cite{Panteleeva:2022uii}
\bibitem{Panteleeva:2022uii}
J.~Y.~Panteleeva, E.~Epelbaum, J.~Gegelia and U.-G.~Mei\ss{}ner,
%``Definition of gravitational local spatial densities for spin-0 and spin-1/2 systems,''
Eur. Phys. J. C \textbf{83}, no.7, 617 (2023),
%doi:10.1140/epjc/s10052-023-11746-x
[arXiv:2211.09596 [hep-ph]].

%\cite{Alharazin:2022xvp}
\bibitem{Alharazin:2022xvp}
H.~Alharazin, B.~D.~Sun, E.~Epelbaum, J.~Gegelia and U.-G.~Mei\ss{}ner,
%``Local spatial densities for composite spin-3/2 systems,''
JHEP \textbf{02}, 163 (2023),
%doi:10.1007/JHEP02(2023)163
[arXiv:2212.11505 [hep-ph]].

%\cite{Panteleeva:2023evj}
\bibitem{Panteleeva:2023evj}
J.~Y.~Panteleeva, E.~Epelbaum, J.~Gegelia and U.-G.~Mei\ss{}ner,
%``Electromagnetic and gravitational local spatial densities for spin-1 systems,''
JHEP \textbf{07}, 237 (2023),
%doi:10.1007/JHEP07(2023)237
[arXiv:2305.01491 [hep-ph]].

%\cite{Chen:2024oxx}
\bibitem{Chen:2024oxx}
Y.~Chen, Y.~Li, C.~Lorc\'e and Q.~Wang,
%``Nucleon axial radius,''
Phys. Rev. D \textbf{110}, no.9, L091503 (2024),
%doi:10.1103/PhysRevD.110.L091503
[arXiv:2405.12943 [hep-ph]].

%\cite{Bernard:2001rs}
\bibitem{Bernard:2001rs}
V.~Bernard, L.~Elouadrhiri and U.-G.~Mei{\ss}ner,
%``Axial structure of the nucleon: Topical Review,''
J. Phys. G \textbf{28} (2002), R1-R35,
%doi:10.1088/0954-3899/28/1/201
[arXiv:hep-ph/0107088 [hep-ph]].
%537 citations counted in INSPIRE as of 25 Nov 2024

\bibitem{Gegelia:1994zz}
J.~Gegelia, G.~S.~Japaridze and K.~S.~Turashvili,
%``Calculation of loop integrals by dimensional counting,''
Theor. Math. Phys. \textbf{101}, 1313-1319 (1994).
% doi:10.1007/BF01018279

\bibitem{Beneke:1997zp}
M.~Beneke and V.~A.~Smirnov,
%``Asymptotic expansion of Feynman integrals near threshold,''
Nucl. Phys. B \textbf{522}, 321-344 (1998).
% doi:10.1016/S0550-3213(98)00138-2
%[arXiv:hep-ph/9711391 [hep-ph]].
  
\bibitem{Hoffmann:2018edo}
S.~E.~Hoffmann,
%``Relativistic probability amplitudes I. Massive particles of any spin,''
[arXiv:1804.00548 [quant-ph]].
% 5 citations counted in INSPIRE as of 02 Dec 2021

%\cite{Weinberg:1995mt}
\bibitem{Weinberg:1995mt} 
  S.~Weinberg,
  ``The Quantum theory of fields. Vol. 1, : Foundations,''
  Cambridge University Press (2005-06-02).
  %%CITATION = INSPIRE-406190;%%
  %169 citations counted in INSPIRE as of 13 Nov 2016

\end{references}
\end{document}